\documentclass[11pt,twoside]{article}
\usepackage{asp2010}
\resetcounters
\begin{document}
\title{Investigating AM Her Cataclysmic Variables
with the Optical Pulsar Timing Analyzer --- OPTIMA}
\author{Aga~S\l{}owikowska,$^1$ Krzysztof~Go\'zdziewski,$^2$
Ilham~Nasiroglu,$^3$ Gottfried~Kanbach,$^4$ Arne~Rau,$^4$
Krzysztof~Krzeszowski$^1$
\affil{$^1$Kepler Institute of Astronomy, University of Zielona G\'ora,
Lubuska 2, 65-265 Zielona G\'ora, Poland}
\affil{$^2$ Toru\'n Centre for Astronomy, Nicolaus Copernicus University, 
Gagarin Str.~11, 87-100 Toru\'n, Poland}
\affil{$^3$ University of \c{C}ukurova, Department of Physics, 01330 Adana, Turkey}
\affil{$^4$Max-Planck Institut f\"ur Extraterrestrische Physik,
Giessenbachstrasse 1, 85741 Garching bei M\"unchen, Germany}}

\begin{abstract}
We focus on short--period eclipsing binaries that belong to a class of
Cataclysmic Variables (CVs). They are known as polars and intermediate
polars, closely resembling their prototype AM Herculis. These binaries
consist of a red dwarf and a strongly magnetic white dwarf, having orbital
periods of only a few hours. Monitoring eclipses of these typically
faint sources demands high-time resolution photometry.  We describe
the very recent results obtained for two CVs, HU~Aqr and DQ~Her, which were
observed with the Optical Pulsar Timing Analyzer (OPTIMA). The new
observations of HU~Aqr confirm that the O--C (Observed minus Calculated)
diagrams exhibit variations known for this binary which can be explained
by a single, massive Jupiter--like planet, possibly accompanied by
a very distant companion.
\end{abstract}
%
%
OPTIMA\footnote{\url{http://www.mpe.mpg.de/OPTIMA}}
is a fast, single-photon sensitive optical photometer and polarimeter
\citep{Straubmeier2001,Kanbach2003, Kanbach2008,Stefanescu2011}.  It uses
optical fibres to gather light from fixed apertures in the focal plane in to
SPAD (Single-Photon Avalanche Diode) detector modules, while the field
surrounding the apertures is imaged using a standard CCD camera.
The photometer part of the instrument contains eight fibre--fed single
photon counters --- SPADs, and
a GPS for the time control.  There are seven fibres in bundle and one
separate fibre located at a distance of 1~arcmin.  Single photons are
recorded simultaneously and separately in all channels with absolute UTC
time--scale tagging accuracy of $\sim 5~\rm{ns}$.  The quantum efficiency of
the SPADs reaches a maximum of 60$\%$ at 750~nm and lies above 20$\%$ in the
450--950~nm range.  

The system was designed from scratch as a guest
instrument, easily adapted to different telescopes.  It can be reconfigured
for photometric, polarimetric or spectroscopic use within one observing run.
OPTIMA was successfully used at various observatories. As its name implied,
OPTIMA was initially designed for optical pulsar studies, however it is not
limited to this subject only. There were many successful measurements acquired
with OPTIMA that are not pulsar related. Some of them are presented here.
%
\section*{DQ~Her Observations by OPTIMA}
%
DQ~Herculis (or Nova Herculis 1934) was a slow, bright nova occurring in
December 1934, reaching a peak magnitude of 1.5 \citep{Adams1935}.  It is
classified as an Intermediate Polar.  The binary consists of a red dwarf star
(M2 type) and a fast rotating, highly magnetic white dwarf (DBe type).  The
rotation period $P\sim72$ seconds is easily resolved in our new OPTIMA
observations.  The orbital period of DQ~Her ($P_{\mbox{\scriptsize
bin}}\sim$~4h~39m) lies above the period gap ($>3$ hours).  DQ~Her
resembles closely the AM Her type stars in many ways, although the
latter are spin--orbit synchronised, and generally have shorter orbital
periods which prevents the formation of an accretion disk.

In \citeyear{Dai2009}, Dai and Qian analysed the available observations of
DQ~Her, spanning almost 50~years.  They reported the presence of a third
object in the system to explain the orbital period variations observed in
the (O--C) diagram.  If the putative third body is confirmed, it would likely
turn out to be a brown dwarf.

\begin{figure}
\includegraphics[width=0.5\linewidth]{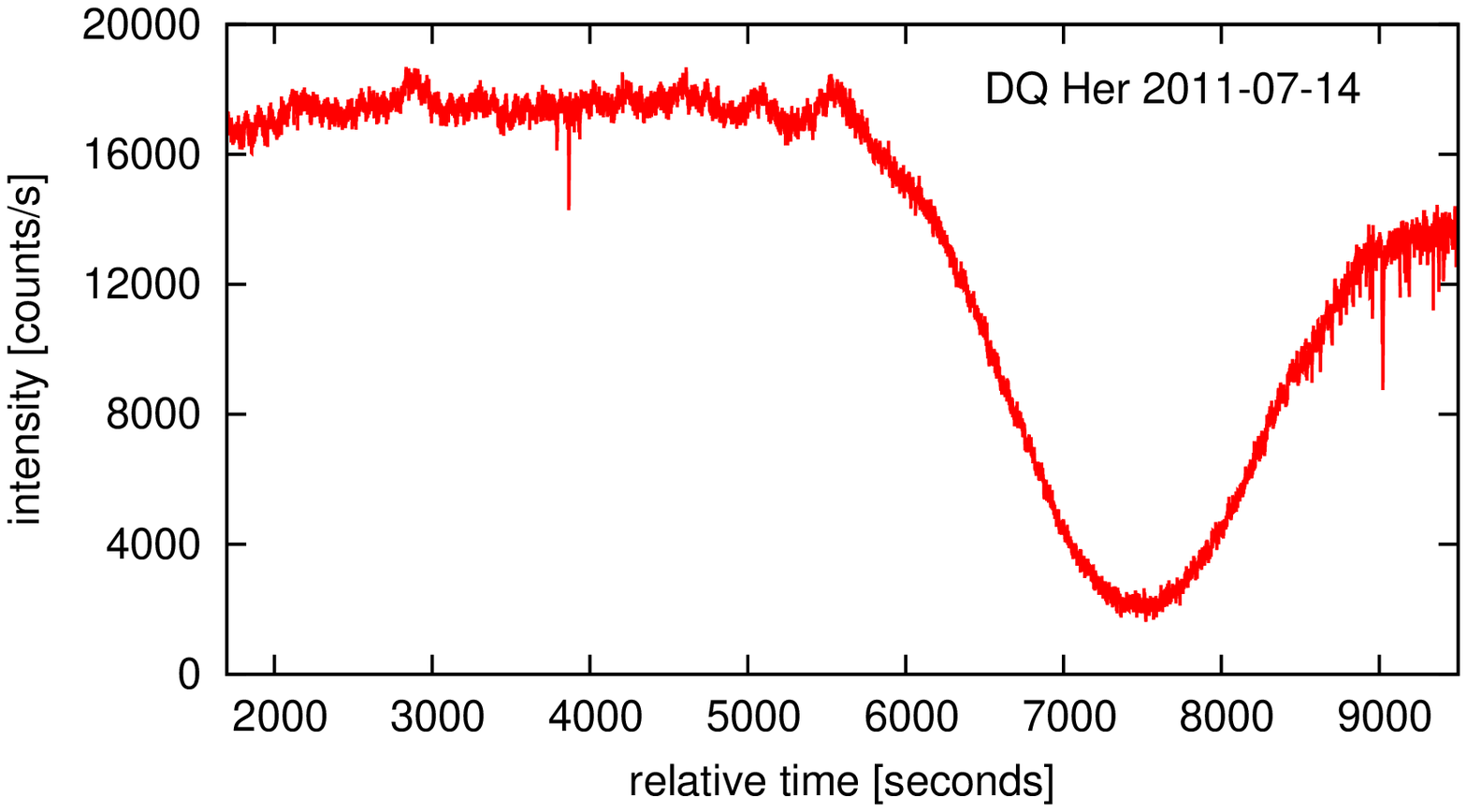}
\includegraphics[width=0.5\linewidth]{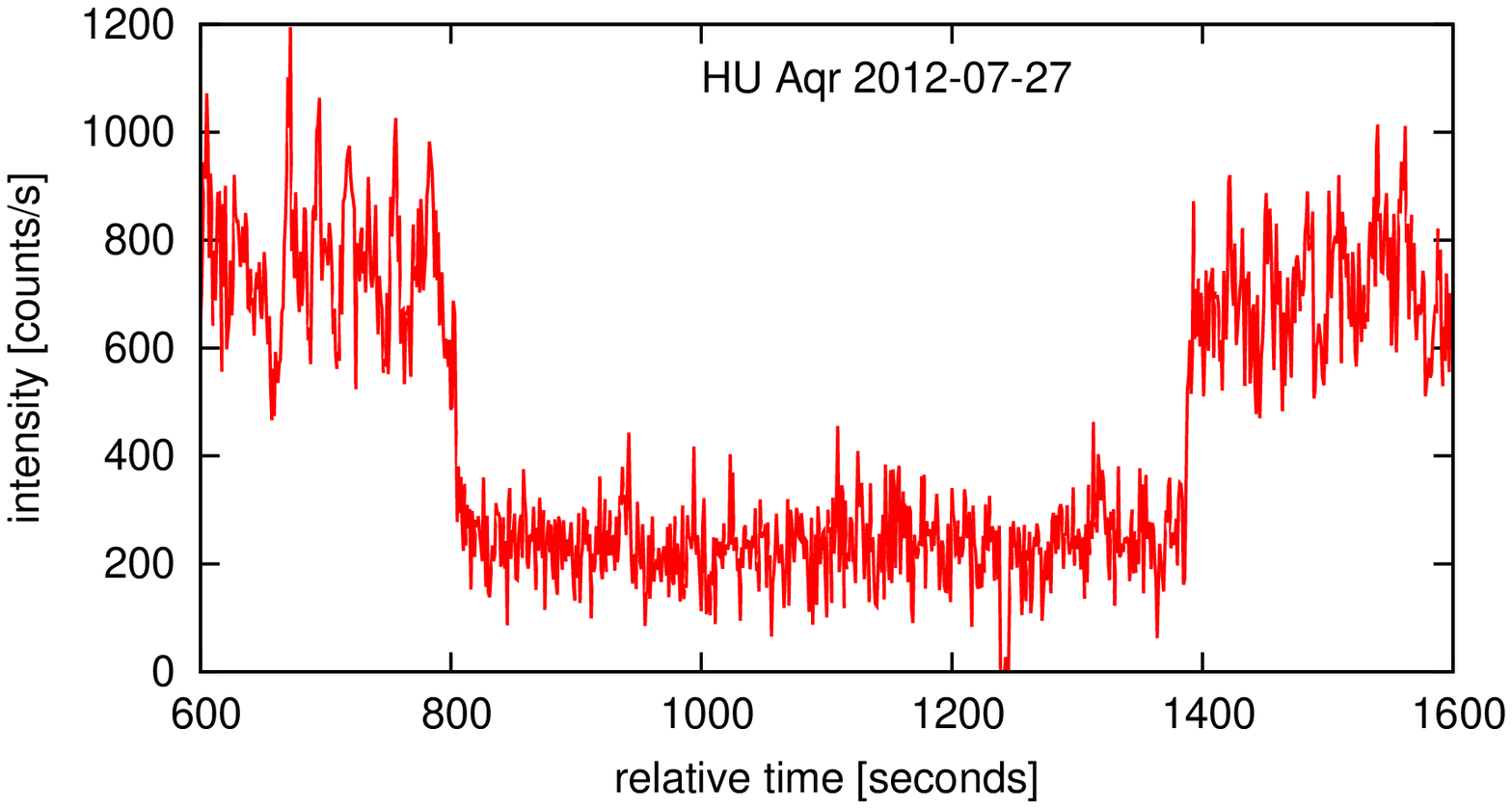}
\caption{
Example light curves of DQ~Her and HU~Aqr obtained during 2011 and 2012
observing campaigns at Skinakas Observatory using the OPTIMA photometer 
working with 1.3-m telescope (the left and the right panel, respectively).
Data binning is 1~second.
}
\label{Fig:lc_both}
\end{figure}

However, the DQ~Her light curves collected over such a long period of time
likely suffer of observational biases and errors.  During the 2011 and 2012
seasons, we gathered new, optical high-time resolution photometric data with
OPTIMA. An example light curve of DQ~Her is shown in the left panel of
Fig.~\ref{Fig:lc_both}.  We expect that such observations will be helpful to
reveal the true origin of the large magnitude (O--C) variations and confirm
or withdraw the planetary hypothesis.  This project requires a long--term
monitoring of the object.

%
%
%
\subsection*{HU~Aqr --- A Single Jovian Companion?}
%
The eclipsing polar HU~Aquarii (HU~Aqr) consists of a strongly magnetic WD
accompanied by a red dwarf (spectral type M4V). The orbital period is 
about 125 minutes. This is one of the brightest polars at X-ray energies and
in the optical domain with visual magnitudes ranging from 14.6 to 18.
Therefore, it has also been one of the most intensively studied systems so far. 

For the sky background monitoring, we usually choose hexagonally located
fibres that are not by chance pointed to any source, hence recording the
background only, and those with the response most similar to that of the
central fibre, when the instrument is targeted at the dark sky.  An example
of the background subtracted light curve, obtained in July 2012, is shown in
the right panel of Fig.~\ref{Fig:lc_both}.

\begin{figure}
\centerline{\includegraphics[width=1.0\linewidth]{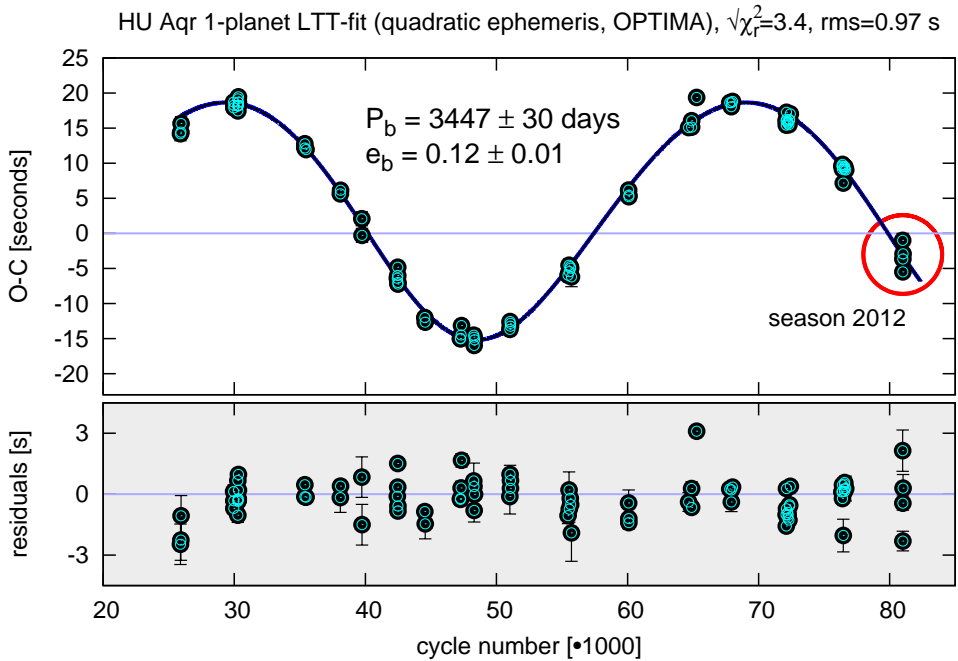}}
\caption{
Synthetic curve of the 1-planet LTT quadratic ephemeris
model to optical OPTIMA measurements (cycles $\sim$~25.000--80.000)
of HU~Aqr.
}
\label{Fig:huaqr2012}
\end{figure}

Since the discovery in 1991, HU~Aqr eclipses are constantly and carefully
monitored.  The polar exhibits relatively large (O--C)
variations unjustified by the astrophysics of the binary. 
\cite{Qian2011} explained this by the presence of two jovian
companions, but the proposed 4-body planetary system was proved strongly unstable
\citep{Horner2012}.  In our recent paper \citep{Gozdziewski2012},
we performed a detailed study of all available data in the literature,
in terms of a new Light Travel Time (LTT) ephemeris model $T_{ep}(l)$,
and carried out new OPTIMA observations.
This model is formulated w.r.t.  the Jacobi coordinates
with the origin in the mass center of the binary, i.e.,
$
\mbox{(O--C)} = T_{ep}(l) - t_0 - l P_{\mbox{\scriptsize bin}}
-\beta l^2 - \sum_p \zeta_p,
$
for a given epoch $t_0$, cycle number $l$, binary period $P_{\mbox{\scriptsize bin}}$
and period damping factor $\beta$ (i.e., the quadratic ephemeris). 
Contribution $\zeta_p(t)$ of a putative planet ($p=1,2,\ldots$) 
to the (O--C) deviation is:
\[
 \zeta_p(t) = K_p \left[ \sin \omega_p \left( \cos E_p(t) - e_p \right) 
      + \cos \omega_p \sqrt{1-e_p^2} \sin E_p(t)
     \right],
\]
where $K_p,e_p,\omega_p$ are the semi-amplitude of the LTT signal,
eccentricity, argument of the pericenter, respectively.
The orbital period
$P_p$ of a companion, and its time of pericenter passage are introduced
indirectly through the eccentric anomaly $E_p(t)$.

We found that the (O--C) variations of HU~Aqr are best explained by a
quasi-periodic signal appearing in terms of the quadratic ephemeris,
which can be interpreted through the presence of $\sim 7$
Jupiter masses planet, in a $\sim 10$~yr quasi-circular orbit
\citep{Gozdziewski2012}.  In fact, this result is achieved thanks to
superior timing accuracy of OPTIMA in the optical domain.  The first OPTIMA
observations of HU~Aqr were performed in 1999.  Since then, further 64
eclipses spanning more than 55.000 orbital cycles have been recorded. Looking at
these observations only, we detected a clear, quasi-sinusoidal variation of
the (O--C).  All available mid-egress moments derived from different
instruments and spectral domains exhibit much more complex and apparently
multi-modal pattern of the (O--C).  It is usually explained in the literature
by the superposition of Keplerian orbits.  Most likely false detection of the second
planet (signal) proposed by \cite{Qian2011} can be related to mixing observation in
different spectral windows and instrumental errors.  This is confirmed by
the best-fit solution to the OPTIMA observations only (Fig.~\ref{Fig:huaqr2012}),
including the very recent data (an example light curve is shown in the right
panel of Fig.~\ref{Fig:lc_both}). It nicely follows our single-planet
(quasi-sinusoidal signal) model with the quadratic ephemeris in
\cite{Gozdziewski2012}.  The origin of relatively
large quadratic term is yet uncertain. It might appear due to very
distant companion.  To confirm or withdraw the planetary hypothesis,
or the (O--C) signal coherence, a long-term monitoring of the HU~Aqr
is necessary.

%
\section*{Conclusions}
%
OPTIMA is an unique instrument, particularly useful to observe short-period
eclipsing CVs binaries. Thanks to its specific design, it provides ultra-high
time resolution photometry and polarimetry, required for characterising
these intriguing objects, and resolving their complex astrophysics.
Recently, these targets become particularly interesting due to a number of
announcements of their planetary companions. OPTIMA observations are crucial
to verify a doubtful discovery of a 2-planet system around HU~Aqr
\citep{Gozdziewski2012}. It might also help to study the LTT effect,
which is presumably present in the case of DQ~Her and other CVs. 

\acknowledgements
AS, IN and KK acknowledge support from the Foundation for Polish Science grant
FNP HOM/2009/11B, as well as from the FP7 Marie Curie European Reintegration
Grant (PERG05-GA-2009-249168). KG is supported by the Polish Ministry of
Science and Higher Education Grant~N/N203/402739 and the POWIEW project
of the European Regional Development Fund in Innovative Economy Programme
POIG.02.03.00-00-018/08. Skinakas Observatory is a collaborative project
of the University of Crete (UoC) and the Foundation for Research and
Technology -- Hellas (FORTH), and the Max-Planck-Institute for
Extraterrestrial Physics. This work was in part supported under
the FP7 Opticon European Network for High Time Resolution Astrophysics
-- HTRA project.
%
\bibliography{slowikowska}

\begin{thebibliography}{}
\expandafter\ifx\csname natexlab\endcsname\relax\def\natexlab#1{#1}\fi
\expandafter\ifx\csname url\endcsname\relax
  \def\url#1{\texttt{#1}}\fi
\expandafter\ifx\csname urlprefix\endcsname\relax\def\urlprefix{URL }\fi
\providecommand{\eprint}[2][]{\url{#2}}

\bibitem[{{Adams} et~al.(1935){Adams}, {Christie}, {Joy}, {Sanford}, \&
  {Wilson}}]{Adams1935}
{Adams}, W.~S., {Christie}, W.~H., {Joy}, A.~H., {Sanford}, R.~F., \& {Wilson},
  O.~C. 1935, \pasp, 47, 205

\bibitem[{{Dai} \& {Qian}(2009)}]{Dai2009}
{Dai}, Z.~B., \& {Qian}, S.~B. 2009, \aap, 503, 883

\bibitem[{{Go{\'z}dziewski} et~al.(2012){Go{\'z}dziewski}, {Nasiroglu},
  {S{\l}owikowska} et~al.}]{Gozdziewski2012}
{Go{\'z}dziewski}, K., {Nasiroglu}, I., {S{\l}owikowska}, A., et~al. 2012,
  \mnras, 425, 930

\bibitem[{{Kanbach} et~al.(2003){Kanbach}, {Kellner}, {Schrey}, {Steinle}
  et~al.}]{Kanbach2003}
{Kanbach}, G., {Kellner}, S., {Schrey}, F.~Z., {Steinle}, H., et~al. 2003, in
  Society of Photo-Optical Instrumentation Engineers (SPIE) Conference Series,
  edited by M.~{Iye}, \& A.~F.~M. {Moorwood}, vol. 4841 of Society of
  Photo-Optical Instrumentation Engineers (SPIE) Conference Series, 82

\bibitem[{{Kanbach} et~al.(2008){Kanbach}, {Stefanescu}, {Duscha},
  {M{\"u}hlegger} et~al.}]{Kanbach2008}
{Kanbach}, G., {Stefanescu}, A., {Duscha}, S., {M{\"u}hlegger}, M., et~al.
  2008, in Astrophysics and Space Science Library, edited by D.~{Phelan},
  O.~{Ryan}, \& A.~{Shearer}, vol. 351 of Astrophysics and Space Science
  Library, 153

\bibitem[{{Qian} et~al.(2011){Qian}, {Liu}, {Liao}, {Li} et~al.}]{Qian2011}
{Qian}, S.-B., {Liu}, L., {Liao}, W.-P., {Li}, L.-J., et~al. 2011, \mnras, 414,
  L16

\bibitem[{{Stefanescu}(2011)}]{Stefanescu2011}
{Stefanescu}, A. 2011, Ph.D. thesis, Technische Universit\"at M\"unchen

\bibitem[{{Straubmeier}(2001)}]{Straubmeier2001}
{Straubmeier}, C. 2001, Ph.D. thesis, Technische Universit\"at M\"unchen

\bibitem[{{Wittenmyer} et~al.(2012){Wittenmyer}, {Horner}, {Marshall},
  {Butters}, \& {Tinney}}]{Horner2012}
{Wittenmyer}, R.~A., {Horner}, J., {Marshall}, J.~P., {Butters}, O.~W., \&
  {Tinney}, C.~G. 2012, \mnras, 419, 3258

\end{thebibliography}
%
\end{document}